%% file: main.tex
\documentclass[acmsmall, nonacm]{acmart}

\usepackage[english]{babel}
\usepackage{blindtext}
\usepackage{multirow}
\usepackage{xcolor}
\usepackage{soul}
\usepackage{array}
\usepackage{tagging}
\usepackage{graphicx}
\usepackage{caption}
\usepackage{cleveref}
\usepackage{subcaption}
\usepackage{adjustbox}
\usepackage[most]{tcolorbox}
\usepackage{float}
\usepackage[inline]{enumitem}
\usepackage{tikz}
\usepackage{xspace}
\usepackage[frozencache,cachedir=.]{minted} 
\usepackage{wrapfig}

\usepackage{makecell}    
\usepackage{booktabs}    

\tcbuselibrary{listings,breakable}

\newcommand{\circled}[1]{%
  \tikz[baseline=(myanchor.base)]%
    \node[circle,fill=black!50,inner sep=1pt] (myanchor)%
      {\color{white}\bfseries\footnotesize #1};}

\AtBeginDocument{%
  }

\acmSubmissionID{\#284}

\begin{document}

\title{A Network Arena for Benchmarking AI Agents on Network Troubleshooting}




\author{Zhihao Wang}
\affiliation{%
  \institution{University of Electronic Science and Technology of China (UESTC)}
  \country{China}
}

\author{Alessandro Cornacchia}
\affiliation{%
  \institution{KAUST}
  \country{Saudi Arabia}
}

\author{Alessio Sacco}
\affiliation{%
  \institution{Politecnico di Torino}
  \country{Italy}
}

\author{Franco Galante}
\affiliation{%
  \institution{Politecnico di Torino}
  \country{Italy}
}

\author{Marco Canini}
\affiliation{%
  \institution{KAUST}
  \country{Saudi Arabia}
}

\author{Dingde Jiang}
\affiliation{%
  \institution{University of Electronic Science and Technology of China (UESTC)}
  \country{China}
}

\renewcommand{\shortauthors}{}

\newcommand{\llm}{LMs\xspace}
\newcommand{\sysname}{\texttt{NIKA}\xspace}
\newcommand{\kathara}{Kathará}


\newif\ifsubmission
\submissionfalse
\submissiontrue

\ifsubmission
    \newcommand{\ac}[1]{}
    \newcommand{\as}[1]{}
    \newcommand{\zw}[1]{}
    \newcommand{\xp}[1]{}
    \newcommand{\mc}[1]{}
\else
    \newcommand{\ac}[1]{\textit{\color{magenta}[Ales: #1]}}
    \newcommand{\as}[1]{\textit{\color{red}[Alessio: #1]}}
    \newcommand{\zw}[1]{\textit{\color{cyan}[Zhihao: #1]}}
    \newcommand{\mc}[1]{\textit{\color{blue}[Marco: #1]}}
    \newcommand{\xp}[1]{\textit{\color{green}[Peng: #1]}}
\fi

\newcommand{\smartparagraph}[1]{\noindent{\bf #1}\ }
\newcommand{\indentparagraph}[1]{{\bf #1}\ }

\crefname{figure}{Fig.}{Figs.}
\crefname{table}{Table}{Tables}
\crefname{section}{Sec.}{Secs.}

\begin{abstract}
Agentic systems, powered by Large Language Models (LLMs), assist network engineers with network configuration synthesis and network troubleshooting tasks. 
For network troubleshooting, progress is hindered by the absence of standardized and accessible benchmarks for evaluating LLM agents in dynamic network settings at low operational effort.
We present \sysname, the largest public benchmark to date for LLM-driven network incident diagnosis and troubleshooting.
\sysname targets both domain experts and especially AI researchers alike, providing zero-effort replay of real-world network scenarios, and establishing well-defined agent-network interfaces for quick agent prototyping.
\sysname comprises hundreds of curated network incidents, spanning five network scenarios, from data centers to ISP networks, and covers 54 representative network issues. 
Lastly, \sysname is modular and extensible by design, offering APIs to facilitate the integration of new network scenarios and failure cases.
We evaluate state-of-the-art LLM agents on \sysname and find that while larger models succeed more often in detecting network issues, they still struggle to localize faults and identify root causes.
\sysname is open-source and available to the community: \texttt{\url{https://github.com/sands-lab/nika}}.
\end{abstract}





\maketitle

\section{INTRODUCTION}

Artificial Intelligence (AI) is shifting from human-oriented chat applications to autonomous agentic systems with decision-making capabilities. 
Google has projected a 1 trillion USD market value for agentic AI by 2040~\cite{google-agentic-ai-report}, with Large Language Models (LLMs) at the core of this transformation.
This evolution has accelerated the adoption of LLMs in domains requiring dynamic, context-aware operations, such as networking~\cite{wu2025netllm,bian,confucius2025,wang2024netassistant, angi2025llnet,wang2024netconfeval,mekrache2024intent,lira2024large, wang2025graph, donadel2024can} and systems~\cite{roy2024exploring,chen2024automatic, yu2024monitorassistant, zhang2024lm,goel2024x,aiopslab,cornacchia2025between,2024rcagent}.
Network troubleshooting is a compelling target: given a network incident, engineers undertake mechanical yet cumbersome steps to identify and remediate issues, from probing the network to explain packet drops, to refining detection logic across heterogeneous telemetry and diagnosis tools~\cite{fayaz2016efficient,arzani2018007}. 
The process is manual, slow, and prone to errors, requiring expert operators to reason across multiple dimensions. Recent work shows that cloud operators and telco vendors have already deployed AI troubleshooting agents for production networks~\cite{bian, confucius2025,wang2024netassistant,ericsson-ana}, leveraging LLMs to parse textual data from monitoring tools and assist operators in failure detection and localization.

Despite the broad consensus, our community still lacks the benchmarks and platforms needed to explore and compare agent designs in a principled way. For example, existing systems such as NetAssistant~\cite{wang2024netassistant}, \textsc{BiAn}~\cite{bian}, and Confucius~\cite{confucius2025} are evaluated in closed settings with no publicly available datasets and limited disclosure of evaluation scenarios. 
NetConfEval~\cite{wang2024netconfeval} instead provides a public benchmark but focuses on static configuration synthesis, making it impossible for live network troubleshooting or studying how well agents sequence decision-making. Additionally, it cannot capture important dimensions that extend beyond accuracy, such as time to resolution, reasoning quality, or the impact on network performance during diagnosis, which are equally critical in troubleshooting tasks.

We can observe two gaps in the existing solutions.
First, designers of AI agents need to evaluate a rich design space that includes, among others:
\emph{agent structure}, \textit{e.g.,} monolithic versus specialized sub-agents~\cite{li2023camel, zhuge2024gptswarm, pan2025why};  
\emph{prompting strategies}, \textit{e.g.,} how to structure the agent's reasoning process~\cite{wei2022chain,wang2023self,madaan2023self,langgraph-workflows}; 
\emph{tool interfaces}, \textit{e.g.,} granularity and abstraction of external APIs~\cite{patil2024gorilla, basu2025nestfulbenchmarkevaluatingllms}; 
\emph{state management}, \textit{e.g.,} which historical data to keep in the context at each reasoning step~\cite{xu2025amem, lee2024human}.
Because AI programmers do not necessarily possess domain knowledge, they expect a ready-to-use benchmark to evaluate the impact of their design choices on agent behaviors at low operational effort.
Second, experimenting with network troubleshooting scenarios requires a controlled network environment in which to connect AI agents and reproduce failures without disrupting normal user traffic and network operations.
Production networks are unsuitable due to safety, privacy, and reproducibility constraints. Existing emulators provide controllable networks, but fall short of offering:
\begin{enumerate*}[label=\emph{(\roman*)}]
  \item streamlined reproducibility of given incident scenarios;
  \item a unified interface to connect AI agents with the network environment;
  \item scoring mechanisms for task objectives (\textit{e.g.,} localization accuracy) and reasoning-level evaluations. 
\end{enumerate*}


To fill these gaps, we present \textbf{\sysname}, a \underline{\textbf{N}}etwork \underline{\textbf{I}}ncidents benchmar\underline{\textbf{K}}ing FrameworK for \underline{\textbf{A}}I Agents.
\sysname is a unified platform that can offer:
\begin{enumerate*}
  \item a benchmark suite of curated network incidents that covers 54 realistic network issues, ranging from link and host failures to resource contention, and includes five network scenarios, four of which can be instantiated at different topology sizes, spanning campus and data center networks. By combining these dimensions, the benchmark yields 640 distinct troubleshooting incidents for evaluating AI agents. The benchmark can be further extended by randomizing failure locations and composing multiple issues within a single incident.
  \item A modular plug-and-play orchestration platform that connects AI agents with the network environment, enabling real-time troubleshooting in realistic conditions, and providing a human-facing interface to monitor agent performance. 
\end{enumerate*}
%

\sysname orchestrates behind the scenes the components required for 
traffic generation, failure injection, integration with specialized telemetry instrumentation and diagnosis tools, observability, performance evaluation, and reasoning trajectory analysis, leaving AI programmers to focus only on agent design and evaluation.
%
Through the Model Context Protocol (MCP)~\cite{mcp}, the framework exposes more than 30 monitoring and troubleshooting tools to third-party agents, ranging from sketches and INT~\cite{int} to SDN controller APIs and switch CLIs.
Thus, we used \sysname to evaluate three popular LLMs, namely GPT-OSS, GPT-5, and GPT-5-mini, and analyzed how these models behave when troubleshooting various network failures. 
We have found that, for example, although larger models are more successful in detecting network issues, they still struggle to localize faults and identify root causes.



\smartparagraph{Contributions.} In summary, this paper makes the following contributions:

\begin{itemize}[leftmargin=*,topsep=0pt]
  \item we formalize the problem of benchmarking AI agents for network troubleshooting, identifying key requirements and challenges;
%
\item we design a structured methodology to benchmark AI agents in this context and implement a modular platform that orchestrates the end-to-end evaluation workflow and outputs detailed observability and performance reports;
\item 
we evaluate three state-of-the-art LLMs to validate the benchmark rigor, summarizing the problems detected with \sysname  and share our experience in dealing with them; 
\item we release the framework~\cite{nika} and disclose a large public dataset of AI agents' behavior for network troubleshooting, with more than 900 reasoning traces.\ac{@zhihao add reference to Zenodo}
\end{itemize}

\section{BACKGROUND \& MOTIVATION}
\label{sec:background}


\smartparagraph{Network troubleshooting.} Network incidents are inevitable with the scale and complexity of current production networks. 
Hardware failures, misconfigurations, and software bugs, are common causes of network incidents~\cite{meza2018large,netseer, bian}.
Because incidents lead to degraded performance and service outages, operators must promptly resolve them and minimize the impact on customers.
Over decades of networking research, a humongous amount of network monitoring and diagnosis tools have been developed to assist operators in troubleshooting incidents in their networks~\cite{arzani2018007, van2018network, dhamdhere2007netdiagnoser,huang2020omnimon,moshref2016trumpet}. In addition to basic networking tools (\textit{e.g.,} \texttt{\small ping}, \texttt{\small traceroute}, etc.), operators leverage advanced diagnostics frameworks such as PingMesh~\cite{guo2015pingmesh}, 007~\cite{arzani2018007}, deTector~\cite{peng2017detector} and others to monitor network health, detect anomalies, and diagnose root causes automatically~\cite{fayaz2016efficient, ben2020pint, geng2019simon,gupta2018sonata,handigol2014know,tan2019netbouncer,weitao2022closedloop}.
Despite these advancements, \emph{manual} intervention is still necessary in the presence of never-seen-before anomalies, or anomalies that sit in the tail percentiles and evade detection. 
To date, large providers report more than 200 incidents per week being resolved manually in production networks~\cite{bian}, or remain unsolved~\cite{meza2018large}.

\begin{wrapfigure}{L}{0.52\columnwidth}
    \centering
    \vspace{-0.5\baselineskip} 
    \includegraphics[width=0.52\columnwidth]{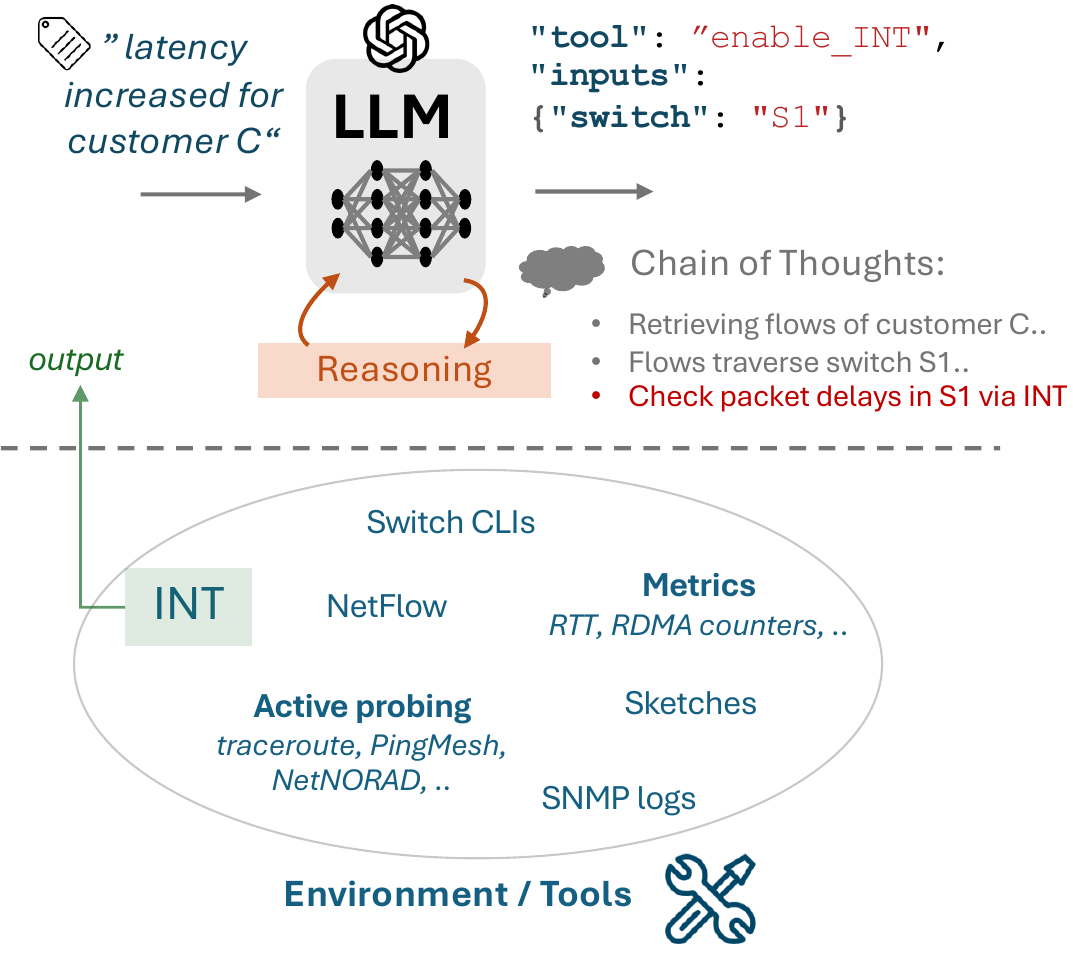}
    \caption{%
      Network troubleshooting with an LLM agent.
    }
    \label{fig:agent-tools}
    \vspace{-\baselineskip} 
\end{wrapfigure}
\smartparagraph{Manual incident management outpaces human capacity.} In a typical troubleshooting workflow, network engineers start receiving tickets containing an incident description. These are high-level descriptions of symptoms reported by customers or automated monitoring systems, \textit{e.g.,} ``increased latency between region A and B'', ``connectivity loss for service S'', etc.
Subsequently, engineers scrutinize multiple alerts generated by their anomaly detection tools and decide which telemetry to inspect in depth.
Examples include logs from network devices, flow-level telemetry (such as NetFlow~\cite{li2013survey} or sFlow~\cite{phaal2001inmon}), and information from data plane telemetry systems (such as INT~\cite{int,ben2020pint,zhao2021lightguardian} or sketches~\cite{sketchovsky, dta,yang2018elastic}). 
Because alerts cardinality scales with the network size and number of monitoring tools, it is common to see hundreds of alerts for a single incident~\cite{zhao2020understanding,kuang2024knowledge}.
Engineers must decide which telemetry to prioritize, reason across multiple signals, and perform manual correlation and causation analysis.
For example, engineers must perform active probing on selected paths or refine the detection logic, \textit{e.g.,} by requesting fine-grained queue-length data to zoom into an ongoing incident~\cite{zoom2net}, enable debug-level diagnostics, etc., until the root cause is identified and remediated.
The process is iterative and time-consuming, admittedly stretching the available human capacity in cloud networks~\cite{bian}.

\textbf{State-of-the-art with AI agents.}
Agentic systems replace or augment human operators with LLM-powered agents 
during the diagnostic workflow~\cite{bian, wang2024netassistant, confucius2025}. 
An LLM-powered \emph{AI agent} combines one or more LLMs with the ability to invoke external \emph{tools} 
-- function calls that let the models interact with the outside world. 
\cref{fig:agent-tools} illustrates the mechanism for a ReAct-style~\cite{yao2023react} reasoning agent.
Given a high-level symptom description such as ``latency has increased for customer C's traffic'', 
the model first performs reasoning steps to map symptoms to potential causes and required telemetry.
\footnote{A \emph{system} prompt instructs the model with the available tools and the agent role (omitted in \cref{fig:agent-tools} for brevity).}
Based on the reasoning steps, the model
generates a tool call, specifying the function to invoke
(\texttt{enable\_INT}) and its parameters (\texttt{switch: S1}). 
The tool executes in the network environment and returns an output, which the model incorporates into subsequent 
reasoning.
%
\footnote{More complex agent architectures explicitly structure the reasoning process as a graph of API calls to LLMs, each with predefined prompt templates and input-output dependencies between calls. Because they follow the same underlying principles, we omit for brevity.}

Large-scale production deployments already validate that LLM-based agents can effectively assist 
network troubleshooting. \textsc{BiAn}~\cite{bian} achieved  95.5\% accuracy 
in localizing faulty devices across 357 incidents at Alibaba Cloud. 
Meta's Confucius~\cite{confucius2025} generalizes this line of work to encompass network management at a large scale.
NetAssistant~\cite{wang2024netassistant} is a copilot for on-call engineers at hyperscaler networks. 
\begin{tcolorbox}[colback=black!5,colframe=black!30,boxrule=0.4pt,top=1pt,bottom=1pt]
\smartparagraph{Evaluation gap.}
However, as the number of relevant applications grows, the establishment of a unified evaluation framework and benchmark has become a critical challenge for the community, which needs to evaluate agent designs consistently and fairly.
Currently, scientific research in this area is fragmented and difficult to compare.  
\end{tcolorbox}

To improve performance and reduce costs, programmers of AI agents must evaluate a rich design space spanning agent architectures, prompting strategies, state management, and tool interfaces.
Yet, no universally accepted benchmark targets network troubleshooting to date, thus creating an evaluation gap for systematic exploration of the problem space.
NetConfEval~\cite{wang2024netconfeval} focuses on configuration synthesis, without closed-loop interaction with a network environment and its dynamically evolving state. 
Systems such as \textsc{BiAn}~\cite{bian} and Confucius~\cite{confucius2025} are evaluated in closed settings with no publicly available datasets, and limited disclosure of the network scenarios. 
It is currently unclear how to advance state-of-the-art, especially for players outside cloud providers or ISPs.
For example, it would be desirable to assess whether \textsc{BiAn}'s design choices (\textit{e.g.,} tool interfaces, prompting strategies, agent structure) generalize to heterogeneous network types, and, if so, how that would impact performance.
Network emulation platforms~\cite{bonofiglio2018kathara, mininet, containernet, containerlab} provide controlled environments and can simulate network incidents.
However, a curated problem set of realistic network incidents is missing. Even when available from operators,
developers would still  
shoulder the effort of reproducing the incidents with high fidelity, \textit{e.g.,} trigger failure events in the presence of given traffic patterns, application-level states and network configurations -- not a trivial task without domain expertise and a time-consuming process with high chances of errors when using low-level emulation APIs.
In addition, these platforms also lack interfaces for agent-network interaction, requiring programmers to build tool-specific adapters to expose the network environment to AI agents. Lastly, there are no standardized evaluation schemes to compare agents' solutions against predefined success criteria, which typically depend on the objective of the troubleshooting task (\cref{sec:nika:benchmark}) at hand.
These limitations either lead programmers of AI agents to build custom 
scenarios for quick prototyping, or dissuade them from experimenting with this research area altogether.

\section{NIKA}\label{sec:nika}

We address the evaluation gap with \sysname.
In this section, we discuss our methodology and its key components.

\subsection{Framework overview}\label{sec:overview}

\begin{figure}[!t]
    \centering
    \includegraphics[width=.97\linewidth]{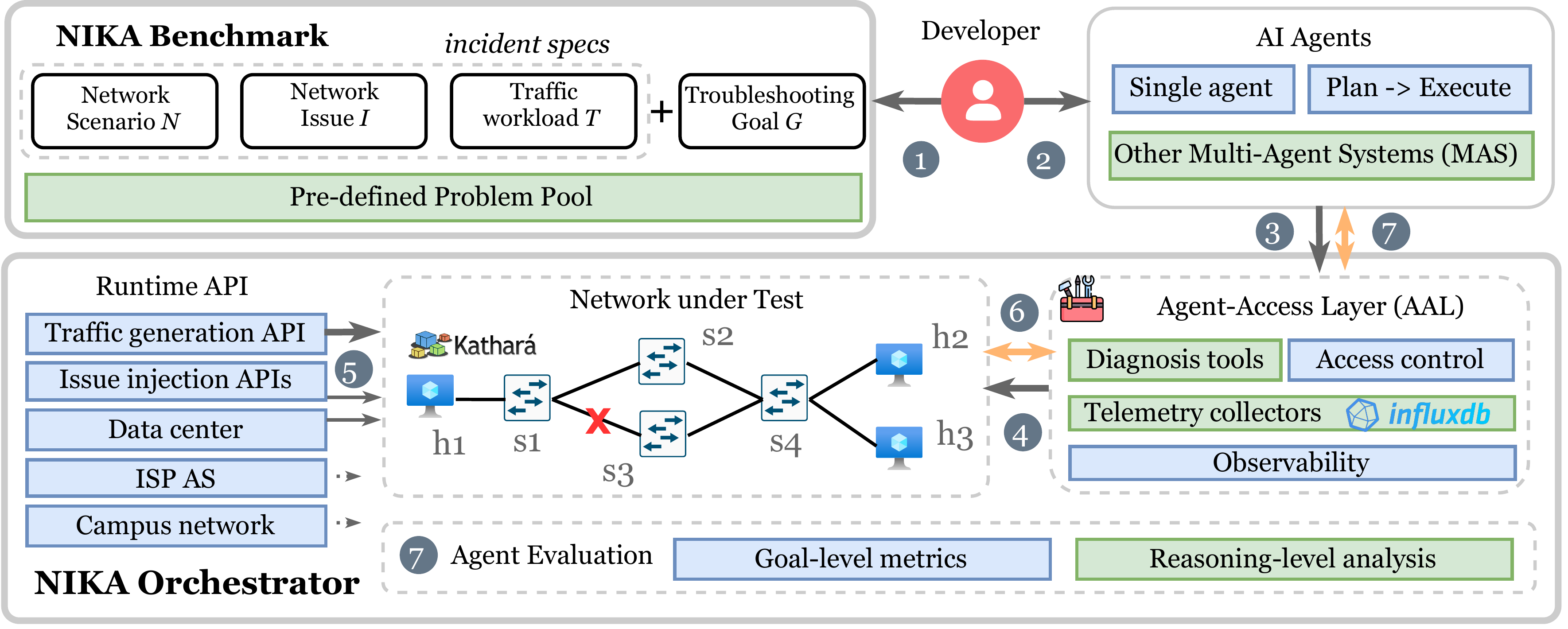}
    \caption{\sysname's architecture. \textbf{Box legend:} \colorbox[HTML]{dae8fb}{blue} $=$ provided by \sysname; \colorbox[HTML]{d5e8d5}{green} $=$ extensible by the developer.}
    \label{fig:architecture}
    \vspace{-1\baselineskip} 
\end{figure}

\sysname is a framework to benchmark troubleshooting AI agents on curated network incidents. 
Its users are developers of AI agents wishing to evaluate a given design. 
As illustrated in \Cref{fig:architecture}, \sysname separates concerns between AI agent development, network environment management, and benchmark creation. 
It consists of two main components: a \emph{benchmark suite} of curated network incidents, and an \emph{orchestration platform} that connects an AI agent with a network environment to reproduce the incident and evaluate the agent's performance.

The developer \circled{1} selects an incident from the benchmark suite (\cref{sec:nika:benchmark}).
Each network incident is defined by a specification of how to reproduce the incident, and a troubleshooting goal for the agent \circled{2}
\textit{e.g.,} localize failures. 
We formalize both in \cref{sec:nika:benchmark:spec}.
Next, the developer interfaces with \sysname's orchestrator (\cref{sec:nika:runtime}) to instantiate the incident in an interactive network-under-test environment and execute the AI agent \circled{3}.
Before agent execution, \sysname's orchestrator initializes the network environment based on the incident specification \circled{4}, \textit{e.g.,} deploys the topology, network configurations, and telemetry instrumentation. 
During agent execution, \sysname exposes \circled{6} the network environment via an \emph{Agent Access Layer} (AAL) with interfaces that the AI agent can invoke. It also orchestrates traffic generation and failure injection \circled{5} to faithfully reproduce the incident in the network environment. 
After agent execution, \sysname evaluates the agent's performance based on goal-specific metrics \circled{7} (\textit{e.g.,} accuracy for root-cause analysis, time to detection for anomaly detection tasks). It also analyzes the agent's reasoning trajectory, \textit{e.g.,} tool usage patterns, to provide insights into the agent's behavior. 



\subsection{Benchmark creation: methodology}
\label{sec:nika:benchmark}

The user selects a network incident from \sysname's benchmark problem pool to evaluate their AI agent.
Each problem in the benchmark is defined by two elements: an \emph{incident specification} that describes how to reproduce the incident in a network environment, and a \emph{troubleshooting specification} that defines the task the agent must perform. 

\subsubsection{Incident specification}\label{sec:nika:benchmark:spec}

The benchmark is a curated pool of network incidents, each referred as \emph{incident specification} and represented as a three-tuple $(\mathcal{N},\mathcal{I},\mathcal{T})$ denoting:

\emph{Network scenario} $\mathcal{N}$: describes 
the network topology, the network devices, protocols and configurations involved. It contains initial configurations and the control plane to initialize all nodes in the network. It also defines which telemetry instrumentation to deploy in the network, \textit{e.g.,} INT, sketches, flow collectors, etc, to restrict the use to user-selected telemetry methods. 
Some legacy clusters may not support advanced switch features such as INT, and users may wish to test how AI agents perform in these clusters. We present the supported network scenarios in~\cref{sec:appendix:scenarios}.

\emph{Network issue} $\mathcal{I}$: defines the network problem. $\mathcal{I}$ serves as the ground-truth evidence against which agent outputs are evaluated. Each issue can be further formalized into $\mathcal{I}=(Dev,Comp,RC)$, where $\textit{Dev}$ is the device with an affected component $\textit{Comp}$ due to a root cause \textit{RC}. 
For example, a link failure issue can be represented as $\mathcal{I}=(\texttt{switch1}, \texttt{eth0}, \texttt{link\_down})$, indicating that the \texttt{eth0} interface on \texttt{switch1} is down due to a link failure. Similarly, a routing misconfiguration issue can be represented as $\mathcal{I}=(\texttt{router2}, \texttt{OSPF}, \texttt{wrong\_area})$, indicating that \texttt{router2}'s OSPF protocol is misconfigured with an incorrect area assignment. \cref{tab:nika:failure_cases} categorizes a selected list of currently supported network issues.

\emph{Traffic workload} $\mathcal{T}$: describes how traffic is generated while reproducing an incident scenario. It includes both regular traffic, which refers to the traffic matrix under normal network operations, and incident-triggering traffic, designed to exacerbate the issue in $\mathcal{I}$. 
For example, a misconfiguration issue may affect regular traffic, yet manifest with more severe symptoms in the presence of specific incident-triggering traffic patterns that deviate from the average traffic matrix, \textit{e.g.,} overload certain network components.
Different combinations of these traffic types allow domain-expert users to modulate troubleshooting complexity by controlling symptom visibility and severity.

\smartparagraph{Incident variations.} Thanks to the incident specification abstraction $(\mathcal{N},\mathcal{I},\mathcal{T})$, \sysname supports systematic variations of the same underlying issue $\mathcal{I}$ by combining different network scenarios $\mathcal{N}$ and traffic workloads $\mathcal{T}$. 
Additionally, because $\mathcal{I}$ itself is modular, the user can generate incident variants by swapping affected devices or root causes, \textit{e.g.,} link failure on different switches, link down \textit{vs.} high error rate. 
This is supported via parametric templates, enabling the generation of practically unlimited incident variants.
Alternatively, the user can create composite incident variations by combining multiple issues $\mathcal{I}$ in the same run, \textit{e.g.,} link failure + routing misconfiguration.

%
%
\begin{wrapfigure}{L}{0.5\columnwidth}
    \vspace{-1.5\baselineskip} 
    \centering
    \begin{subfigure}{0.5\columnwidth}
      \centering
      \input{assets/algo-nikaproblem.tex}
    \end{subfigure}
    \vfill
    \vspace{\baselineskip}
    \begin{subfigure}{0.5\columnwidth}
      \centering
      \input{assets/agent.tex}
    \end{subfigure}
    \caption{\sysname APIs to define and instantiate an incident.\label{fig:codes}}
\end{wrapfigure}

\subsubsection{Troubleshooting specification}\label{sec:nika:benchmark:goal}
Along with the incident specification, each problem in the \sysname's problem pool is associated with a \emph{troubleshooting goal} $\mathcal{G}$. 
The AI agent is tasked with $\mathcal{G}$ for the problem at hand. 
Following the taxonomy used in prior works on cloud diagnosis~\cite{aiopslab}, we adopt a hierarchical structure to define troubleshooting goals, consisting of three levels of increasing complexity: \emph{detection}, \emph{localization}, and \emph{root cause analysis} (RCA). 
\emph{Detection} is a binary decision of the agent, \textit{i.e.,} normal or anomalous network state. Such a detection task is relevant even in production systems that raise alerts for anomalies to disambiguate false positives and false negatives (which are inevitable).  
\emph{Localization} identifies the responsible components $\mathcal{I}_{Dev}$ and $\mathcal{I}_{Comp}$ as a multi-class selection task. \emph{RCA} additionally suggests the underlying root causes, $\mathcal{I}_{RC}$. 
Each level maps to goal-specific evaluators to assess agent performance (\cref{sec:nika:runtime}).

\smartparagraph{Putting-together example.}
\Cref{algo:incident-definition} illustrates how we define the problems in \sysname's problem pool. 
In this example, the network scenario $\mathcal{N}$ is a data center network (ln.\ref{line:scenario}), and the network issue $\mathcal{I}$ is  a resource contention problem due to incast traffic (ln.\ref{line:inject}). 
It accepts the affected device and interface as input parameters (ln.\ref{line:init}), enabling incident variants. 
The traffic workload $\mathcal{T}$ consists of uniform traffic (ln.\ref{line:regulartraffic}). 
The default troubleshooting goal $\mathcal{G}$ is to detect that incast occurred. 
\sysname mandates that the core logic to reproduce the incident is encapsulated in the \texttt{run} method (ln.\ref{line:run})
to manage the incident lifecycle. 
This hook will be invoked by \sysname orchestrator (\cref{sec:nika:runtime}) once the user connects an AI agent and starts the experiment.
Lastly, the problem definition heavily hinges on \sysname's runtime APIs--- ln.\ref{line:scenario},\ref{line:odflowgen}, and \ref{line:regulartraffic}, corresponding to \emph{runtime APIs} in \cref{fig:architecture} ---which we describe next.

\subsection{Orchestrator}\label{sec:nika:runtime}
The orchestrator \emph{consumes} the input incident specification $(\mathcal{N}, \mathcal{I}, \mathcal{T})$, and \emph{materializes} it into execution in a network environment. 
We walk through the example in \cref{algo:nika-agent} to illustrate how \sysname achieves this in practice.

\subsubsection{Network environment and Runtime APIs}
The user selects and instantiates an incident from the benchmark, with concrete parameters for $\mathcal{N}$ (ln.\ref{line:instantiateproblem} in \cref{algo:nika-agent}).
The incident instance is then registered with the \sysname's orchestrator (ln.\ref{line:registernika}). 
When the user connects an AI agent (ln.\ref{line:agent}) and starts the experiment (ln.\ref{line:runbench}), internally the orchestrator invokes the incident's \texttt{run()} method (ln.\ref{line:run} in \cref{algo:incident-definition}) to reproduce it.
\sysname adopts emulation as a default execution backend because it prioritizes fidelity.
However, \sysname's runtime APIs provide high-level abstractions to hide low-level emulation details from the user. 
For example, \sysname exposes APIs to manage common network topologies (ln.\ref{line:scenario}), deploy and instrument measurement tools in the network nodes (ln.\ref{line:deploy}), and generate traffic matrices (ln.\ref{line:odflowgen}). It also provides modules to inject network issues. The issue injection (\cref{fig:architecture}) APIs leverage Linux Traffic Control (TC) for link-level issues, \texttt{stress-ng} for software contentions, and custom scripts for device-level failures such as process crashes and misconfigurations.

\subsubsection{Agent Access Layer (AAL)}\label{sec:nika:runtime:aal}
The Agent Access Layer serves as an access gateway to the network environment for the AI agents. It abstracts the following functionalities.

\emph{1) Telemetry collectors \& diagnosis tools}; AI agents notoriously benefit from structured and well-defined APIs rather than raw CLIs usage and telemetry processing
Methods and interfaces available to human engineers (such as dashboards and telemetry backends) are not necessarily ready-to-use by LLM agents.
AAL implements well-defined interfaces with structured input-output formats that agents can invoke to observe the network state and perform diagnosis actions. 
More specifically, \sysname is already equipped with more than 30 MCP tools for active measurements (\textit{e.g.,} ping probing, raw packet dump), passive telemetry (\textit{e.g.,} counter reads, flow queries, INT records and sketches), and telemetry retrieval (\textit{e.g.,} InfluxDB~\cite{ahmad2017hands}). The detailed list is in ~\cref{tab:nika:mcp_tools}.

\emph{2) Agent access policies}; 
\sysname restricts the action scope of AI agents via user-defined polices. The policies reflect network boundaries and a given ownership model.
For instance, in a data center scenario, the user may want tenant-side agents to be limited to actions within the tenant's boundary, whereas operator-scoped agents act on the entire fabric.
\sysname's policies can be specified in a declarative way
(ln.\ref{algo:accesspolicies}~in~\cref{algo:nika-agent}), and enforce access control at runtime via the AAL (ln.\ref{line:registernika}~in~\cref{algo:nika-agent}). 

\emph{3) Observability}. 
The AAL intercepts every tool call at the boundary between the agent and the network, and provides observability for both tool usage and
network state when the agent queries it. For tool usage, the AAL logs each tool invocation with its input parameters, timestamp, and response. 
For network state, the AAL snapshots the ground-truth telemetry state at the time of each tool invocation, allowing users to inspect agent reasoning in the context of the actual network conditions. Snapshots are persisted beyond the experiment 
lifetime to support post-mortem querying and analysis. 
When multiple agents (\textit{e.g.,} tenant-scoped and operator-scoped) interact with the same network, the AAL provides a unified view of all tool invocations. 
\sysname adopts OTel~\cite{otel} standardized formats to integrate AAL observability with agent reasoning traces, collected by the agent's frameworks.

\input{assets/benchmark_short.tex}

\subsubsection{Evaluator}
Finally, \sysname evaluates the agent's performance based on the troubleshooting goal $\mathcal{G}$.
\sysname implements goal-specific evaluators for detection, localization, and RCA tasks, and for each of them compares agent outputs against the ground truth $\mathcal{I}$ defined in the incident specification.
For localization and RCA, both agent's output and ground truth are represented as boolean masks with
entries indicating the presence or absence of faulty devices or root causes. From these lists, the evaluator computes the confusion matrix and then derives the accuracy.
Beyond accuracy, \sysname also supports several common efficiency metrics, such as time to detection, token usage, and number of tool invocations. 
Finally, \sysname tracks flow latency and packet loss during the diagnosis process to quantify side effects on network performance, and reports violations to user-defined SLOs.

\section{EXPERIMENTAL RESULTS}
\label{sec:evaluation}

\subsection{Setup}
\smartparagraph{\sysname implementation.}
We build \sysname on top of \kathara~\cite{bonofiglio2018kathara}, an open-source container-based network emulator. 
Traffic workloads are generated via \texttt{iperf3} and \texttt{ApacheBench}, and
%
the MCP-based AAL is implemented using FastMCP~\cite{fastmcp} servers.
Finally, we collect all agent reasoning trajectories in LangSmith~\cite{langsmith} and a local datastore, including tool invocations, LLM prompts, and responses. In total, we release 900 execution traces alongside \sysname.

\smartparagraph{Agent implementation.}
We adopt a two-step agentic workflow: the first step executes the troubleshooting tasks and produces an analysis report. The second step then extracts the outputs required to \sysname's evaluator and submits them by invoking the \texttt{eval} callback in \cref{algo:incident-definition}.
Our prompts are shared in ~\cref{sec:appendix:prompts}.
All agents follow the ReAct paradigm~\cite{yao2023react} and are implemented in LangGraph. 
We experiment with two private backend LLMs, GPT-5-mini and GPT-5, via the official OpenAI API, and with a third open-source LLM, GPT-OSS:20B~\cite{gpt-oss}, hosted on an NVIDIA GeForce RTX 4090 GPU with 24 GB of VRAM.

\subsection{Performance results}

From the full benchmark detailed in~\cref{sec:appendix:issues}, we construct an evaluation subset of 150 incidents. Specifically, we instantiate each issue in one of its typical scenarios. 
We evaluate three backend models, running each model twice with randomly selected faulty devices ($\mathcal{I}.\textit{Dev}$) and components ($\mathcal{I}.\textit{Comp}$) per run. Under this setup, executing the full evaluation requires approximately 7 hours for GPT-OSS:20B, 10 hours for GPT-5-mini and 15 hours for GPT-5 on an Ubuntu 22.04 VM with 16 CPU cores and 32~GB of RAM.

\smartparagraph{Even with SOTA LLMs, troubleshooting is challenging without specialized agent designs.}
We report the results in \Cref{tab:eval:llm}.
As expected, GPT-5 consistently outperforms alternatives across all troubleshooting objectives, improving detection w.r.t. GPT-5-mini by roughly 20\%, localization by nearly $2\times$, and RCA accuracy by more than $2.5\times$. These improvements are even more pronounced when compared against GPT-OSS:20B.
Besides better accuracy, GPT-5 exhibits qualitatively stronger reasoning behavior: it issues more tool invocations while requiring fewer reasoning steps, consumes fewer input tokens (105k \textit{vs.} 156k), and produces substantially more output tokens (14.6k \textit{vs.} 6.5k) and reasoning budget (12.3k \textit{vs.} 4.5k tokens).
These improvements come at a higher average per-incident runtime, which increases by $\approx50\%$\ when moving from GPT-5-mini to GPT-5. 

\begin{figure*}[t]
    \centering
    \includegraphics[width=0.4\linewidth]{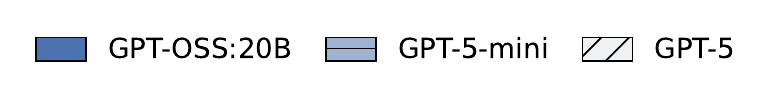}
    \vfill
    \vspace{-1ex}
    \begin{subfigure}{0.31\linewidth}
        \centering
        \includegraphics[width=\linewidth]{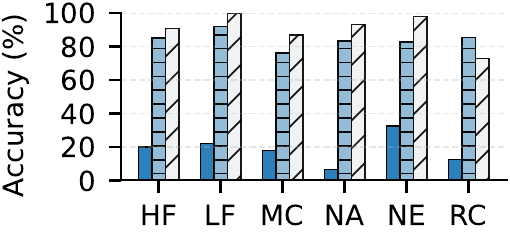}
        \caption{Detection}
        \label{fig:cate:det_acc}
    \end{subfigure}
    \begin{subfigure}{0.31\linewidth}
        \centering
        \includegraphics[width=\linewidth]{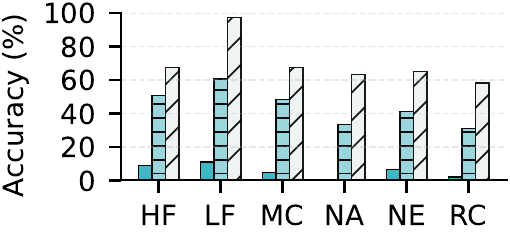}
        \caption{Localization}
        \label{fig:cate:loc_acc}
    \end{subfigure}
    \begin{subfigure}{0.31\linewidth}
        \centering
        \includegraphics[width=\linewidth]{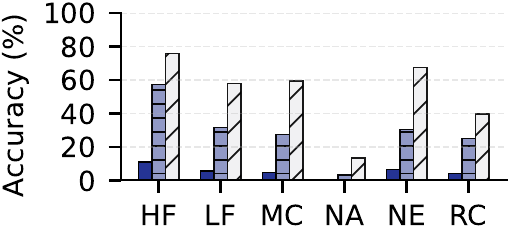}
        \caption{RCA}
        \label{fig:cate:rca_acc}
    \end{subfigure}
    \vspace{-1\baselineskip}
    \caption{GPT-5 agent versus network issues type (\cref{tab:nika:failure_cases}). \textbf{Legend}: 
    \emph{\textbf{LF}}: Link Failure, 
    \emph{\textbf{NE:}} Network node Error, 
    \emph{\textbf{NA:}} Network under Attack, 
    \emph{\textbf{EF:}} End-host Failure, 
    \emph{\textbf{MC:}} MisConfiguration, 
    \emph{\textbf{RC:}} Resource Contention.
    }
    \label{fig:eval:category}
\end{figure*}

However, even for GPT-5, the overall performance remains far from perfect.
To better understand the limitations of current LLMs for network troubleshooting, we analyze the performance of the three LLMs across different issue categories in \cref{fig:eval:category}.
GPT-5 maintains high detection accuracy across the majority of issue categories, with resource contention as the main exception.
Localization and RCA exhibit an even larger variation: link failure (LF) achieves the highest localization accuracy (97\%), while resource contention remains challenging (58\%). 
We manually investigated the agent reasoning traces and network telemetry, recorded by \sysname's observability module, and observed that for incidents with less direct symptoms, the agent tends to stop at shallow, connectivity-centric explanations. For example, in resource contention-related problems (\textit{e.g.,} \texttt{\small link\_packet\_corruption}, \texttt{\small NF\_load\_balancer\_overload}, \texttt{\small sender\_resource\_contention}), despite persistent indicators such as CRC errors, queue drops, and asymmetric losses, the agent settles on vague diagnoses such as ``link unstable'' or ``transient congestion'', without connecting these symptoms to the deeper causes.

\begin{table}[!t]
\caption{Performance on \sysname benchmark for different LLMs (columns are averages).}
\vspace{-1\baselineskip}
\centering
\resizebox{\textwidth}{!}{
\scriptsize
\begin{tabular}{ccccccccccc}
\toprule
\textbf{Model} & \textbf{Time (s)} & \textbf{\# Steps} & \textbf{\# Tools} & \textbf{\# In tokens} & \textbf{\# Out tokens} & \textbf{\# Rea. Tokens} & \textbf{\# Det. Acc.} & \textbf{\# Loc. Acc.} & \textbf{\# RCA Acc.} \\
\midrule
GPT-OSS:20B & 175.8 & 12.6 & 11.7 & 69267.1 & 7645.3 & - & 19.0\% & 5.5\% & 5.5\% \\ 
GPT-5-mini & 242.6 & 9.5 & 15.0 & 156050.1 & 6578.6 & 4532.7 & 74.0\% & 36.0\% & 22.0\% \\ 
GPT-5 & 359.2 & 7.0 & 28.4 & 105171.9 & 14603.0 & 12352.4 & 89.0\% & 68.7\% & 55.3\% \\ 
\bottomrule
\end{tabular}
}
\label{tab:eval:llm}
\vspace{-1\baselineskip}
\end{table}

\begin{figure*}[t]
    \centering
    \begin{subfigure}{0.49\linewidth}
        \centering
        \includegraphics[width=\linewidth]{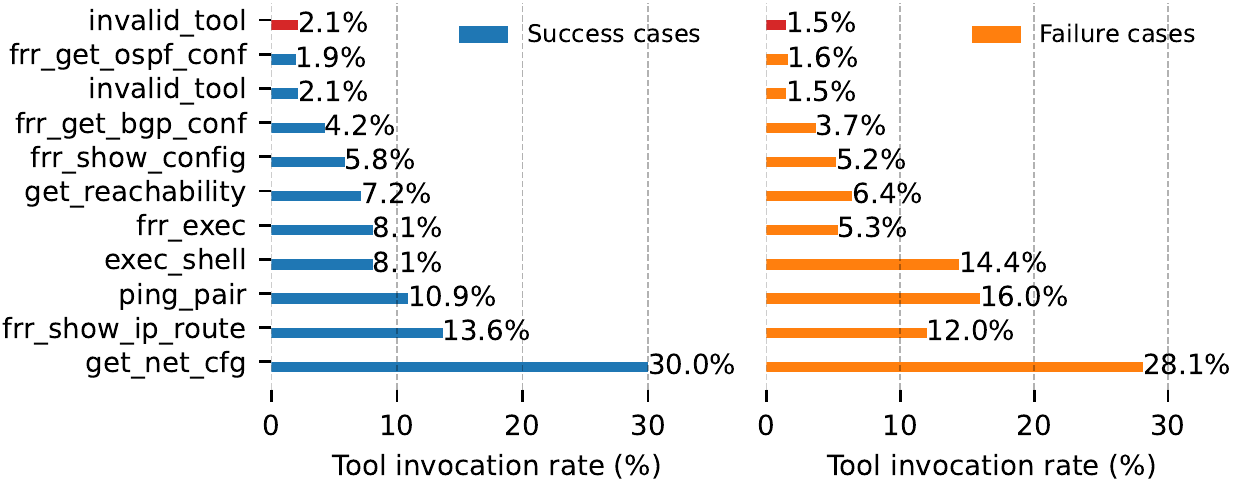}
        \caption{GPT-5-mini}
        \label{fig:tool_gpt_mini}
    \end{subfigure}
    \begin{subfigure}{0.49\linewidth}
        \centering
        \includegraphics[width=\linewidth]{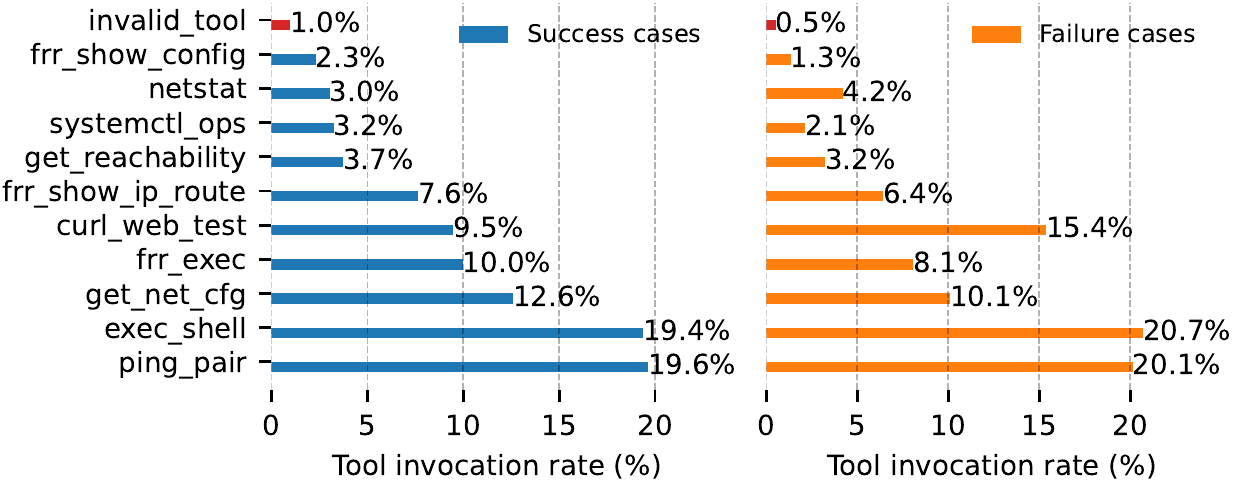}
        \caption{GPT-5}
        \label{fig:tool_gpt}
    \end{subfigure}
    \caption{Tool invocation distribution comparison between successful and failed troubleshooting.
    }
    \label{fig:tool-usage}
    \vspace{-1\baselineskip}
\end{figure*}

\smartparagraph{Tool usage analysis.}
Additionally, we analyze tool invocation patterns to understand inefficiencies.
\cref{fig:tool-usage} compares GPT-5-mini and GPT-5 across success cases (all goals $\mathcal{G}$ achieved) and failure cases (at least one goal failed).
%
In the successful cases, GPT-5-mini (\cref{fig:tool_gpt_mini}) relies mostly on high-level checks, \textit{e.g.,} \texttt{\small get\_host\_net\_config}, remarking that the model can manage troubleshooting at the network layer. GPT-5 (\cref{fig:tool_gpt}), in contrast, makes broader use of application-layer diagnostics, such as \texttt{\small exec\_shell} and \texttt{\small curl\_web\_test}, reflecting its ability to seek out issues across multiple layers.
Given that many issues in our benchmark affect application performance without fully breaking connectivity, GPT-5-mini's limited use of higher-layer tools directly contributes to its weaker diagnostic performance. 
Lastly, this application-layer diagnostic appears more frequently in GPT-5-mini’s failed runs, suggesting that while the model recognizes its relevance, it fails to incorporate it effectively into its reasoning process.




To quantify erroneous tool invocations, we capture and count the exceptions raised during tool calls, including parameter validation errors at the MCP client and execution failures at the MCP server. As shown by the top red bars in \cref{fig:tool-usage}, both models exhibit remarkably low error rates, averaging 1.6\% for GPT-5-mini and 0.7\% for GPT-5 across all success and failure cases. These errors are typically minor parameter mismatches, which contrasts with the significant tool hallucination issues reported in prior work~\cite{cornacchia2025between,aiopslab}, underscoring the importance of structured and well-documented MCP-based interfaces in the AAL.

\begin{figure*}[t]
    \centering
    \includegraphics[width=0.4\linewidth]{figures/plots/model_legend.pdf}
    \vfill
    \vspace{-1ex}
    \begin{subfigure}{0.19\linewidth}
        \centering
        \includegraphics[width=\linewidth]{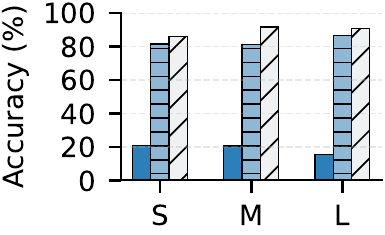}
        \caption{Detection}
        \label{fig:size:det_acc}
    \end{subfigure}
    \begin{subfigure}{0.19\linewidth}
        \centering
        \includegraphics[width=\linewidth]{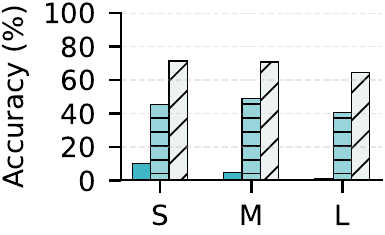}
        \caption{Localization}
        \label{fig:size:loc_acc}
    \end{subfigure}
    \begin{subfigure}{0.19\linewidth}
        \centering
        \includegraphics[width=\linewidth]{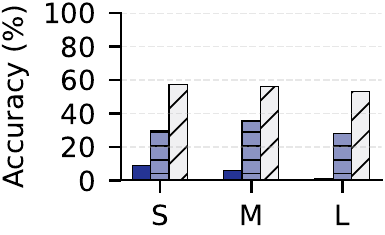}
        \caption{RCA}
        \label{fig:size:rca_acc}
    \end{subfigure}
    \begin{subfigure}{0.19\linewidth}
        \centering
        \includegraphics[width=\linewidth]{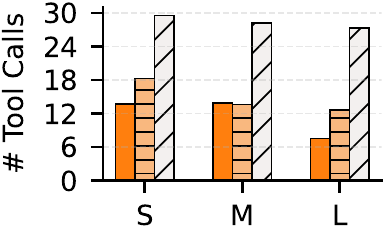}
        \caption{\# Tool Calls}
        \label{fig:size:tool}
    \end{subfigure}
    \begin{subfigure}{0.19\linewidth}
        \centering
        \includegraphics[width=\linewidth]{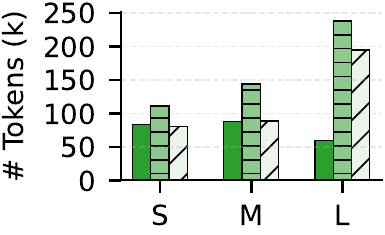}
        \caption{\# Tokens}
        \label{fig:size:token}
    \end{subfigure}
    
    \caption{Agent scalability versus network topology size.}
    \label{fig:eval:size}
    \vspace{-0.5\baselineskip}
\end{figure*}

\smartparagraph{Sensitivity to topology size.}
Finally, we evaluate the impact of network size by using the \sysname runtime API to instantiate small (S), medium (M), and large (L) topologies for four scenarios: data center, campus, ISP backbone, and SDN-enabled cloud POP fabric. On average, these topologies contain 11, 27, and 101 nodes, respectively.
\cref{fig:eval:size} summarizes our findings.
Detection accuracy remains relatively stable across scales, indicating that anomalous symptoms remain visible as networks grow. In contrast, localization and RCA degrade with size, reflecting the ambiguity introduced by larger and more interconnected environments. Although tool invocations decrease slightly with scale, total token consumption nearly doubles for GPT-5-mini and GPT-5, suggesting that larger topologies force agents to process substantially more context, making precise fault isolation increasingly difficult.

\section{DISCUSSION \& LIMITATIONS}\label{sec:discussion}

We surface potential use cases and highlight areas for improvements.


\smallskip
\smartparagraph{\sysname as an educational platform.}
The rise of LLMs is transforming networking education, with faculty grappling with questions about responsible AI use and how to foster deep understanding in an era of AI-assisted learning. 
\sysname can support networking education by exposing students to systematic troubleshooting through a curated incident repository and AI-generated reasoning traces. Instructors can design assignments where students analyze agent behavior, compare their own diagnostics against AI outputs, and identify failure modes. This enables teaching both networking fundamentals and the limitations of current LLM-based systems. We publicly release 900+ reasoning traces to facilitate this use case.

\smallskip
\smartparagraph{Network backend.}
\sysname inherits the limits of network emulation: it cannot faithfully reproduce issues that manifest only in high-speed networks~\cite{10.1145/3131365.3131375}, or that require specialized hardware. 
However, we believe that a significant number of issues can still be meaningfully studied in emulated environments by scaling down link speeds, traffic loads, and the time resolution of network measurements, while preserving the core causal relationships and the fidelity of troubleshooting tools. 
Nevertheless, \sysname's architecture itself remains environment-agnostic and only assumes that a network backend can reproduce $(\mathcal{N}, \mathcal{I}, \mathcal{T})$ and expose tools through a common interface, so alternative backends (\textit{e.g.,} simulators with realistic adapters) are feasible but left to future work. 

\smallskip
\smartparagraph{Mitigation and safe agent actions.}
\sysname currently focuses on diagnosis tasks (detection, localization, RCA) but does not yet support evaluating mitigation actions. Enabling this capability presents non-trivial challenges: evaluating whether an agent-proposed remediation is correct requires both checking syntactic correctness and, crucially, assessing whether its execution would be safe and non-disruptive to the network. To address this, we plan to integrate Batfish~\cite{batfish} as an additional tool in the AAL, enabling agents to validate candidate remediation plans against the network model before applying them. This will allow future versions of \sysname to provide automated verification of mitigation strategies without risking the emulated network.

\section{RELATED WORK}
\label{sec:related}

\smartparagraph{Network monitoring.}
A long line of systems has advanced telemetry and diagnosis for large-scale networks~\cite{guo2015pingmesh, arzani2018007, dhamdhere2007netdiagnoser, van2018network,moshref2016trumpet,huang2020omnimon, gupta2018sonata,handigol2014know,geng2019simon,peng2017detector,tan2019netbouncer,ben2020pint,weitao2022closedloop}.
These works provide rich measurement substrates and partial automation for operators, whereas \sysname targets the complementary problem of autonomous decision-making troubleshooting agents that must choose tools, interpret heterogeneous telemetry, and synthesize diagnoses.

\smartparagraph{LLMs for network operations.}
Microsoft first outlined the vision of AI-driven network incident management for the Azure cloud~\cite{hamadanian2023holistic}.
\textsc{BiAn}~\cite{bian} assists operators at Alibaba Cloud in localizing faulty devices across large-scale WANs using telemetry and historical incident data.
Confucius~\cite{confucius2025} generalizes this idea to multi-agent LLM-based intent-driven management for Meta's networks, while NetAssistant~\cite{wang2024netassistant} targets on-call support for ByteDance.
These systems validate the feasibility of LLM-powered network copilots in production, but do not offer an open, reusable benchmark and execution environment for troubleshooting agents.

\smartparagraph{Benchmarks for LLMs in networking.}
NetConfEval~\cite{wang2024netconfeval} and NetLLMBench~\cite{netllmbench} are public benchmark for network configuration synthesis from natural language intents. They focus on static configuration tasks, whereas \sysname targets dynamic troubleshooting in live network environments.
NetPress~\cite{zhou2025netpress}, similar to \sysname, proposes dynamically generated LLM benchmarks for network applications, however it focuses on network management at large, while \sysname specifically targets troubleshooting tasks. Moreover, it lacks an end-to-end, reusable toolchain that enables accessible experimentation for a broad research and practitioner community. By contrast, \sysname provides APIs for several network scenarios, fault injection, tool integration, traffic generation and observability.

\smartparagraph{LLM research for DevOps.} Beyond networking, LLMs have been used to analyze telemetry data~\cite{zhang2025promassistantleveraginglargelanguage, seshagiri2024chattinglogsexploratorystudy, 2025chatts, llm-trace, liu2024largelanguagemodelsdeliver, alnegheimish2024largelanguagemodelszeroshot}, perform root-cause analysis in cloud systems~\cite{aiopslab,chen2024automatic,roy2024exploring,kashef,yu2024monitorassistant,llexus-fonseca}, and assist with software engineering~\cite{xia2025demystifying, xu2025openrca,chen2024teaching, chen2022codet, hong2024metagpt}. Network troubleshooting shares similarities with these domains, and we expect \sysname to facilitate cross-pollination of ideas and techniques.

\section{CONCLUSION}\label{sec:conclusion}
We designed \sysname, a unified benchmarking framework for AI-driven network troubleshooting. \sysname offers a curated suite of incident scenarios covering a broad spectrum of real-world faults and network topologies, along with an orchestration platform that connects agents to network environments and exposes their behavior for evaluation.
It requires minimal user involvement beyond selecting a network incident from the benchmark suite and providing the AI agent implementation, yet its modular architecture allows advanced users to customize and extend each component.
We see \sysname as a first step in accelerating innovation in AI-driven network diagnosis, and we hope for community contributions to expand it to cover more scenarios and network types.



\bibliographystyle{ACM-Reference-Format}
\bibliography{references}  

\newpage
\appendix
\label{sec:appendix}
\input{appendix.tex}

\end{document}

%% file: assets/algo-nikaproblem.tex
    \begin{tcolorbox}[
        colback=gray!5,
        colframe=black!70,
        boxrule=0.5pt,
        arc=2pt,
        left=2pt,
        right=2pt,
        top=2pt,
        bottom=2pt,
        boxsep=0pt
    ]
    \begin{minted}[
        linenos=true,
        fontsize=\scriptsize,
        numbersep=3pt,
        frame=none,
        framesep=0mm,
        baselinestretch=1.0,
        escapeinside=||
    ]{python}
 class IncidentBase:
     prompt_detection = "You are a network engineer.."
     def run(self):
        self.net.deploy().instrument("default")|\label{line:deploy}|
     # ..   
 class DataCenterIncast(IncidentBase):
     name = "single_link_datacenter_incast"
     def __init__(self, dev, intf, dcn_size="s"):|\label{line:init}|
        self.network = DataCenterClos(size=dc_size)|\label{line:scenario}|
        # .. network issue parameters
     def inject_incident(self):|\label{line:inject}|
         od = ODFlowGenerator(self.network)|\label{line:odflowgen}|
         od.set_src_dst(src="all", dst=dev)
         od.start_traffic(interval=20, speed="X Mbps")|\label{line:injectend}|
     def run(self):|\label{line:run}|
         UniformTraffic(self.network, rho=.4).start()|\label{line:regulartraffic}|
         self.inject_incident()
     def eval(self, answer, goal="detect"):|\label{line:eval}|
         Evaluator().eval(self, answer, goal)|\label{line:evalend}|
    \end{minted}
    \end{tcolorbox}
    \caption{Problem definition in \sysname's benchmark.\label{algo:incident-definition}}

%% file: assets/agent.tex
\centering
    \begin{tcolorbox}[
        colback=gray!5,
        colframe=black!70,
        boxrule=0.5pt,
        arc=2pt,
        left=2pt,
        right=2pt,
        top=2pt,
        bottom=2pt,
        boxsep=0pt
    ]
    \begin{minted}[
        linenos=true,
        fontsize=\scriptsize,
        numbersep=3pt,
        frame=none,
        framesep=0mm,
        baselinestretch=1.0,
        escapeinside=||
    ]{python}
 from nika.benchmark import DataCenterIncast
 from nika.runtime import Orchestrator
 from langchain.agents import create_agent
 access_policy = { |\label{algo:accesspolicies}|
     "nodes": ["pod0.switch*"]; 
     "tools" ["*"]
 }
 prob = DataCenterIncast("s1", "eth0") |\label{line:instantiateproblem}|
 NIKA = Orchestrator(env=prob, aal_cfg=access_policy)|\label{line:registernika}|
 tools = NIKA.get_tools()
 agent = create_agent("gpt-5", tools=tools) |\label{line:agent}|
 NIKA.benchmark(agent) |\label{line:runbench}|
\end{minted}
\end{tcolorbox}
\caption{Instance creation and agent onboarding.\label{algo:nika-agent}}

%% file: assets/benchmark_short.tex
\begin{table}[t!]
\caption{Selected network issues ($\mathcal{I}$) in the \sysname benchmark. Full incident list is in~\Cref{tab:nika:failure_cases_full}.}
\vspace{-1\baselineskip}
\centering
\resizebox{\textwidth}{!}{
\scriptsize
\begin{tabular}{@{}p{0.2\textwidth}p{0.1\textwidth}p{0.28\textwidth}p{0.4\textwidth}}
\toprule
\textbf{Category}  & \textbf{\# Issues} & \textbf{Representative issue} & \textbf{Key Signals} \\
\midrule
Link failures & 6 & Link flap & flap event logs; packet drops \\

End-host failures  & 10 & Conflicting VPN memberships &  Overlapping subnets; VPN servers unreachable\\

Network node errors  & 8 & Number of MPLS labels hit limit & Error logs; packet drops \\

Misconfigurations  & 14 & BGP ASN mismatch & BGP session fails; ASN mismatch detected \\

Resource contention  & 6 & Microbursts on interface  & Reduced throughput; queue buildup \\

Network under attack & 10 & Service DoS & Surge in HTTP connections; CPU/RAM usage spikes \\
\bottomrule
\end{tabular}
}
\label{tab:nika:failure_cases}
\vspace{-0.5\baselineskip}
\end{table}

%% file: appendix.tex
\section{\sysname's incident pool}

\subsection{Network issues}
\label{sec:appendix:issues}
\sysname benchmark consists of a broad and spectrum of realistic network issues, detailed in ~\Cref{tab:nika:failure_cases_full}.
\input{assets/benchmark.tex}

\subsection{Network scenarios}
\label{sec:appendix:scenarios}
\sysname provides a suite of canonical, ready-to-use topologies in~\Cref{tab:nika:network_scenarios}.

\begin{table}[h!]
\centering
\caption{Network scenarios supported in the \sysname benchmark.}
\label{tab:nika:network_scenarios}
\resizebox{\textwidth}{!}{
\scriptsize
\begin{tabular}{@{}p{0.3\textwidth}p{0.05\textwidth}p{0.55\textwidth}@{}}
\toprule
\textbf{Scenario} & \textbf{Scalable} & \textbf{Description} \\
\midrule
Data center network (CLOS) & \checkmark &
Multi-tier leaf--spine fabric with edge servers. \\
Campus network (3-tier) & \checkmark &
Enterprise core--distribution--access topology. \\
ISP backbone network (meshed) & \checkmark &
Provider-style backbone with core and access nodes. \\
SDN-enabled cloud POP fabric (CLOS/star) & \checkmark &
SDN fabric with centralized controller and edge switches. \\
P4 programmable testbed & -- &
Compact testbed for data-plane algorithms and pipeline validation. \\
\bottomrule
\end{tabular}
}
\end{table}

\section{\sysname runtime}

\subsection{Supported MCP tools in \sysname}
\sysname equips more than 30 troubleshooting MCP tools. \cref{tab:nika:mcp_tools} presents a selected list.

\begin{table}[h!]
\centering
\caption{Selected MCP tools in \sysname.}
\label{tab:nika:mcp_tools}
\resizebox{\textwidth}{!}{
\scriptsize
\begin{tabular}{@{}p{0.18\textwidth}p{0.2\textwidth}p{0.5\textwidth}@{}}
\toprule
\textbf{Category} & \textbf{Tool} & \textbf{Description} \\
\midrule
\multirow{7}{=}{Active measurements} 
& Get reachability & Check pairwise reachability among all hosts \\
& Ping & ICMP echo for reachability and latency \\
& Traceroute & Trace packet forwarding paths \\
& Iperf & Measure achievable end-to-end bandwidth \\
& Http latency probe & Measure HTTP request/response latency \\
& TCP connect test & Test TCP connection establishment to a target via \texttt{nc} \\
& Execute CLI (dual) & Execute commands on two devices simultaneously \\
\midrule
\multirow{6}{=}{Passive measurements} 
& Port counters & Retrieve interface statistics (bytes, packets, errors) \\
& Statistics query & Query device statistics (e.g., TC, \texttt{ethtool}) \\
& Flow table query & Inspect forwarding or flow-table entries \\
& Routing table query & Retrieve routing information from routers \\
& Log retrieval & Fetch system and event logs \\
& Config retrieval & Fetch device configuration files \\
\midrule
\multirow{2}{=}{Telemetry retrieval} 
& InfluxDB query & Query time-series telemetry data \\
& Sketch query & Query traffic sketches and summaries \\
\bottomrule
\end{tabular}
}
\end{table}

\subsection{Agents' prompts used in \sysname}
\label{sec:appendix:prompts}

\begin{tcolorbox}[
  title={Prompt used for diagnosis agent},
  colback=gray!3,
  colframe=gray!50,
  fonttitle=\bfseries,
  breakable
]
\begin{Verbatim}[fontsize=\small, breaklines=true, breakanywhere=true, breaksymbol={}]
You are a network troubleshooting expert.
Focus on:
(1) detecting if there is an anomaly; (2) localizing the faulty devices; (3) identifying the root cause.
Basic requirements:
- Use the provided tools to gather necessary information.
- Do not provide mitigation unless explicitly required.
\end{Verbatim}
\end{tcolorbox}

\begin{tcolorbox}[
  title={Prompt used for submission agent},
  colback=gray!3,
  colframe=gray!50,
  fonttitle=\bfseries,
  breakable
]
\begin{Verbatim}[fontsize=\small, breaklines=true, breakanywhere=true, breaksymbol={}]
You are an expert network engineer.
Your task is to submit the final solution for this network problem based on the diagnosis reported provided.
Carefully review the diagnosis results and ensure that your submission is accurate and complete.
You must strictly follow the submission format and call the submit() tool to submit your solution.
\end{Verbatim}
\end{tcolorbox}

%% file: assets/benchmark.tex
\begin{table}[t!]
\centering
\caption{Network issues ($\mathcal{I}$) and related incidents in the \sysname benchmark.}
\resizebox{\textwidth}{!}{
\scriptsize
\begin{tabular}{@{}p{0.15\textwidth}p{0.28\textwidth}p{0.45\textwidth}r@{}}
\toprule
\textbf{Category} & \textbf{Root Cause} & \textbf{Key Signals} & \textbf{\# Incident} \\
\midrule

\multirow{6}{0.25\textwidth}{Link failures} 
 & Link flap & flap event logs; packet drops & 26 \\
& Link detached & Physical link not detected; PHY down & 26 \\
 & Link down & Interface state down & 26 \\
 & Faulty cable & CRC errors; corrupted frames & 26 \\
 & MAC address conflict & Same MAC seen on multiple ports; MAC flapping logs & 26\\
 & Link fragmentation disabled & Large packets dropped; MTU mismatch & 26 \\
\midrule

\multirow{6}{0.25\textwidth}{End-host failures} 
& Conflicting VPN memberships &  Overlapping subnets; VPN servers unreachable & 3 \\
& Host crash & Host unresponsive; no heartbeat; ping fails & 35 \\ 
& Host IP conflict & Duplicate IP alerts; ARP conflict detected & 26 \\
& Host IP misconfig & Incorrect or missing IP address; host unresponsive & 68 \\ 
& Incorrect netmask & Partial reachability; inconsistent routing behavior & 16 \\
& DNS empty answer & Incorrect or missing DNS records; NXDOMAIN & 6 \\ 
 \midrule

\multirow{5}{0.25\textwidth}{Network node errors}
& Number of MPLS labels hit limit & Error logs; packet drops & 1 \\
& Switch/router crash (e.g., overheating) & Switch down and unreachable from MGMT & 20 \\ 
 & P4 program reads \texttt{invalid} header field & Packet drops; error logs (plaform-dependent) & 8 \\ 
 & SDN controller crash & Switches isolated; new flows dropped & 6 \\
 & Southbound port unreachable & OpenFlow/TCP 6633/6653 unreachable & 12 \\ 
 \midrule

\multirow{11}{0.25\textwidth}{Misconfigurations\\(routing, ACL, etc.)} & BGP ASN mismatch & BGP session fails; ASN mismatch detected & 7 \\

 & BGP blackhole route leak & Traffic to specific prefixes blackholed; unexpected AS path & 7 \\
 & Missing BGP advertisement & Prefix not propagated; missing announcements & 7 \\
 & Host static blackhole & Static blackhole route active; traffic dropped & 7 \\
 & OSPF area misconfiguration & OSPF adjacency failure; area mismatch & 6 \\
 & OSPF neighbor missing & Missing neighbor; no Hello packets exchanged & 6 \\
 & Forwarding table entry misconfig & No matching entry: default drop & 8 \\ 
& Flow rule loop & Traffic loop observed; CPU spike; port flooding & 6 \\
 & Flow rule shadowing & Lower-priority rule overridden by higher-priority rule & 6 \\
 & ARP ACL block & ARP requests or replies dropped; ACL deny counters increase & 26 \\
 & ICMP ACL block & ICMP traffic blocked; ping failes & 26 \\
 & Routing control-plane ACL block & BGP (TCP/179) or OSPF (IP proto 89) blocked; neighborship fails & 13 \\ 
 & HTTP ACL block & HTTP 80/443 traffic blocked; client connection timeout & 12 \\
 \midrule

 \multirow{5}{0.25\textwidth}{Resource contention} & Microbursts on interface  & Reduced throughput; queue buildup & 26 \\
& Receiver saturated \& slow & Multiple segments ACKed per ACK, RWND $<$ CWND; & 12 \\
 & Incast traffic & Queue buildup; packet drops; retransmissions & 12 \\
 & Sender saturated \& slow & Segments smaller than MSS; Flight size $<$ min(CWND,RWND) & 24 \\ 
 & Software middle-box overloads & CPU usage saturates; queue buildup; RTT increases & 3 \\
 \midrule

\multirow{6}{0.25\textwidth}{Network under attack} & Service DoS & Surge in HTTP connections; CPU/RAM usage spikes & 18 \\ 

 & BGP hijacking & More specific or illegitimate prefixes appear; path anomaly 
 & 3 \\

& DHCP spoofing & DHCP clients received spoofed configurations (IP, DNS, etc.)
 & 9 \\ 
 & DNS spoofing & DNS points to wrong addresses
 & 12 \\ 
 & ARP cache poisoning & Abnormal traffic redirection & 26 \\
 & Misaligned sketch thresholds & Triggers false-positive cardinality alerts (e.g., DoS), packet drops
 & 1 \\
\midrule
\textbf{Total} & - & - & 640 \\
\bottomrule
\end{tabular}
}
\label{tab:nika:failure_cases_full}
\end{table}